\documentclass[
    reprint,
amsmath,amssymb,
aps,
]{revtex4-2}

\usepackage{graphicx}

\begin{document}

\title{Effects of correlated collisions and intermittency on the growth of lucky droplets}

\author{Tobias B\"atge}
\affiliation{Max Planck Institute for Dynamics and Self-Organization, (MPI DS), Am Faßberg 17, 37077 G\"ottingen, Germany}
\affiliation{Faculty of Physics, University of G\"ottingen, Friedrich-Hund-Platz 1, 37077 G\"ottingen, Germany}

\author{Johannes Zierenberg}
\affiliation{Max Planck Institute for Dynamics and Self-Organization, (MPI DS), Am Faßberg 17, 37077 G\"ottingen, Germany}
\affiliation{Faculty of Physics, University of G\"ottingen, Friedrich-Hund-Platz 1, 37077 G\"ottingen, Germany}

\author{Michael Wilczek}
\affiliation{Theoretical Physics I, University of Bayreuth, Universitätsstr. 30, 95447 Bayreuth, Germany}
\affiliation{Max Planck Institute for Dynamics and Self-Organization, (MPI DS), Am Faßberg 17, 37077 G\"ottingen, Germany}
\email[To whom correspondence should be addressed.]{Michael.Wilczek@uni-bayreuth.de}

\begin{abstract} 
To trigger precipitation, water droplets in warm clouds need to attain a sufficient size. 
Theoretical estimates based on condensation and gravitational collisions alone fail to explain the observed timescales for the onset of precipitation for a range of droplet sizes. 
This suggests the involvement of collisional growth mediated by turbulence to resolve the so-called ``size-gap problem''. 
For the onset of rain, it is sufficient that statistical outliers, coined ``lucky droplets'', cross the size gap. 
In this study, we explore the influence of turbulence on droplet growth, focusing on correlated collisions and intermittency. 
Using direct numerical simulations of droplets in turbulent flow, we constrain a non-Markovian stochastic framework that allows us to assess memory effects on the droplet-size distribution arising from correlations between consecutive collisions.
Using our framework, we find that correlated collisions accelerate the initial growth of lucky droplets but have sub-leading effect at later stages. 
Consequently, we neglect correlations from collisions and model an ensemble of cloud parcels
representing fluctuations in the volume-averaged dissipation rate. 
Here, the distribution of droplet sizes in each parcel is described by a linear master equation with a time-dependent collision rate according to the volume-averaged dissipation rate.
Our analyses of this toy model show that intermittency can significantly reduce the time required by lucky droplets to cross the size gap.
\end{abstract}

\maketitle

In warm clouds, the so-called ``size gap'' constitutes a major open problem in the formation of rain droplets~\cite{Shaw2003, Grabowski2013}.
While droplets smaller than $15\mu\text{m}$ spontaneously form due to condensation~\cite{wallace2006atmospheric}, and droplets larger than $50\mu\text{m}$ rapidly grow due to gravity-induced differential settling~\cite{Pumir2016}, it remains a subject of investigation how droplets can grow from 15 - 50$\mu\text{m}$ and bridge the size gap within the typical timescale for rain onset of about $30\text{min}$ \cite{Grabowski2013}.

Here, turbulence is believed to be vital by amplifying growth through collisions~\cite{Devenish2012}.
It may introduce spatial clustering~\cite{sundaram_collins_1997,Zaichik2009, petersen2019} and high relative velocities~\cite{Ayala2008, pan2010}, where also the so-called sling effect~\cite{Falkovich2002, Bewley2013} contributes to the collision kernel~\cite{Falkovich2002, wilkinson2006caustic,Vosskuhle2014}.
However, even with turbulence the average time between collisions of small droplets is typically on the order of hours and thus too long to explain rain onset~\cite{Pumir2016}.

An elegant explanation of rapid rain formation despite slow average collision events is through statistical fluctuations~\cite{Telford, Twomey1966}. 
As collisions are a random process, there is a small probability to observe the rare event of fast-growing droplets that were initially called ``fortunate"~\cite{Telford} and later coined ``lucky droplets"~\cite{Kostinski2005}.
Previous results on these rare events~\cite{Kostinski2005, Wilkinson2016, wilkinson2023quantifying} imply that lucky growth in turbulence could be a crucial ingredient to explain rain initiation.

However, these estimates assume collision rates to be constant with time, while collisions occur preferably in distinct regions of the flow~\cite{Picardo2019} and successive inertial particle collisions are known to be correlated~\cite{sundaram_collins_1997, Bec2016}. 
By means of a generalized collision-coalescence framework, such correlated collisions have been shown to accelerate particle growth \cite{Bec2016}.
Additionally, spatio-temporal fluctuations of the volume-averaged dissipation rate render collision rates time dependent as droplets pass through regions of varying dissipation rates~\cite{perrin2014preferred}. 
Both aspects could further accelerate the growth of lucky droplets, and we here aim to investigate their effect under representative turbulence conditions in warm clouds.

Cloud conditions involve high-Reynolds-number turbulence that comes with a large scale separation between the largest and smallest spatio-temporal scales in the flow along with strong intermittency. 
Intermittency manifests itself in extreme spatio-temporal fluctuations of the dissipation rate.
The statistics of these fluctuations can be captured by various modeling approaches such as the refined similarity hypothesis (K62)~\cite{Oboukhov1962, Kolmogorov1962}, the $\beta$-model~\cite{frisch1978}, as well as multifractal models~\cite{Benzi1984, MENEVEAU1987, Frisch1995}, a corresponding large-deviation formulation~\cite{Fouxon2020}, or a model which accounts for observed deviations of log-normality~\cite{Elsinga2020}.
These fluctuations are often accounted for only on the scales attainable by simulations. However, with the extreme range of scales present in clouds, those fluctuations are more pronounced and might also lead to even more pronounced fluctuations in droplet growth and thus more extreme statistical outliers, i.e., lucky-droplets.

In this study, we investigate the growth of lucky droplets in a turbulent flow and incorporate spatio-temporal fluctuations in the volume-averaged dissipation rate through an ensemble model.
Using direct numerical simulations of cloud parcels with fixed volume-averaged dissipation, we find that strong dissipation can induce correlated collisions that may accelerate droplet growth on short timescales but turn out to be a sub-leading correction at later times. 
This is consistent with previous reasoning~\cite{Kostinski2005, Wilkinson2016, wilkinson2023quantifying}.
Based on our numerical results, we parameterize a simple toy model of collision growth in a cloud parcel with a 
time-dependent volume-averaged dissipation rate to model temporal fluctuations.  
Subsequently, we evaluate collision growth in an ensemble of cloud parcels, modeling spatio-temporal fluctuations, to find that such fluctuations of the volume-averaged dissipation can substantially accelerate the formation of lucky droplets.

\begin{figure}
    \centering
    \includegraphics[width=1\columnwidth]{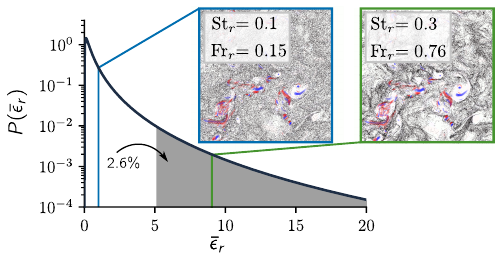}
    \caption{
    Intermittency in clouds causes strong fluctuations of the normalized dissipation rate $\overline{\varepsilon}_r$ averaged over a volume of extent $r$, here modeled as a lognormally distributed (see text for details).
    As a result, parameters that depend on the average dissipation rate, such as the Stokes number of droplets of a given size and the Froude number, vary locally across a cloud. 
    We can mimic this in numerical simulations by varying these parameters, see insets, which show
    typical simulation snapshots for $\epsilon_r=\langle\epsilon\rangle$ (blue frame) and $\epsilon_r=9\langle\epsilon\rangle$ (green frame), where particles are in black, regions of strong vorticity in blue, and regions of strong strain in red.
       }
    \label{fig:upper_per}
\end{figure}

\section*{Model and Methods}

We model a parcel of cloud turbulence by direct numerical simulations (DNS) of Navier-Stokes turbulence. 
While turbulence in clouds spans multiple scales, from the integral scale $L\sim 100\,\text{m}$ down to the Kolmogorov length scale $\eta_K\sim 1\,\text{mm}$~\cite{Pruppacher2010}, our simulations are limited to length scales of $\mathcal{O}\left(1\,\text{m}\right)$ and times scales of $\mathcal{O}\left(1\text{s}\right)$ where we approximate the flow by homogeneous isotropic turbulence. 
We consider the DNS box as a representation of a small cloud parcel, see, e.g., \cite{Saito2018, thomas2020diffusional,chen2020impact, Abade2025}. 
The key idea for the following analysis is to take into account dissipation fluctuations on scales larger than the individual simulation box through ensemble modeling. 
This approach offers a complementary perspective to Large Eddy Simulations, where the large scales are explicitly resolved and the small scales are modeled.

On the length scale of our simulation box, we estimate the amplitude of dissipation fluctuations with the refined similarity hypothesis~\cite{Kolmogorov1962, Oboukhov1962} that was shown to reasonably describe intermittent features in in-situ measurements, experiments, and simulations~\cite{SIEBERT2010426,lawson2019direct}.
In this case, the volume-averaged dissipation rate $\epsilon_r$ averaged over a sphere of radius $r$ (or box of side length $r$ in our case) is log-normally distributed~\cite{Kolmogorov1962, Oboukhov1962}: 

\begin{align}
    P(\bar\epsilon_r)=\frac{1}{\bar\epsilon_r \sigma\sqrt{2 \pi}}\exp{\left(-\frac{(\ln \bar\epsilon_r+\sigma^2/2)^2}{2\sigma^2}\right)}\,,
    \label{eq:lognorm}
\end{align}
where $\bar\epsilon_r=\epsilon_r/\langle\epsilon\rangle$ is the averaged dissipation rate rescaled by the global mean, and the variance is determined by the length-scale ratio $L/r$,

\begin{align}
    \sigma^2=\ln A +\mu\ln\frac{L}{r}\quad\text{with}\quad \langle\bar\epsilon_r^2\rangle=A\left(\frac{L}{r}\right)^\mu.
\end{align}
Here, $\mu\approx0.25$~\cite{Pope2000} and $A$ is an order-one constant that depends on the large scales and, in the following, is set to $A=1$ for simplicity. 
Note that the log-normal distribution of the dissipation rate - while being a reasonable model approximation - is not a fully accurate description as there are deviations from the log-normal distribution, especially for high Reynolds numbers, see e.g. \cite{Elsinga2020}.  
The heavy-tailed nature of $P(\bar\epsilon_r)$ implies that simulation parcels can have a very high volume-averaged dissipation rate (Fig.~\ref{fig:upper_per}). 
For the example of an integral scale $L = 100\,\text{m}$ and a cloud parcel on the scale of $r=0.25\,\text{m}$ (this corresponds to our simulation box for a high $\epsilon_r$, see SI for more details on the simulation setup), we find that the 2.6\% most dissipative cloud parcels have on average a volume-averaged dissipation rate that is 9 times larger than the global mean ($\bar\epsilon_r=1$).

Within the turbulent flow of each simulation, we model droplets as spherical Stokes particles.
The droplet position $\boldsymbol{x}(t)$ and velocity $\boldsymbol{v}(t)$ change according to Stokes drag and gravity~\cite{Maxey1983}

\begin{align}
    \dot{\boldsymbol{x}}(t)=\boldsymbol{v}(t) \qquad\dot{\boldsymbol{v}}(t)=\frac{1}{\tau_d}[\boldsymbol{u}(\boldsymbol{x}(t),t)-\boldsymbol{v}(t)]+g\boldsymbol{e}_z\,,
\end{align}
where $\boldsymbol{u}$ denotes the turbulent flow velocity, $g$ the gravitational acceleration and $\tau_d=\frac{2}{9}\frac{a^2}{\nu}\frac{\rho_\text{d}}{\rho_\text{f}}$ is the particle response time.
Here, $a$ denotes the droplet radius, $\nu$ the kinematic viscosity, $\rho_\text{d}$, and $\rho_\text{f}$ denote the density of the droplet and surrounding fluid, respectively. 
This neglects terms from added mass or history forces, which become small due to a high density ratio between water and air as well as the small droplet radius compared to the Kolmogorov length, cf.~\cite{Maxey1983, Pumir2016}.
If two droplets collide, they merge, i.e., we assume a collision efficiency of one; see SI for more details on the collision procedure. 

The system of the turbulent flow and Stokes particles is characterized by three parameters:
the Stokes number $\text{St}=\tau_d/\tau_K$ quantifies the relevance of inertial effects. 
Here, $\tau_K =\sqrt{\nu/\langle \epsilon \rangle}$ is the Kolmogorov time determined from the mean dissipation rate and the kinematic viscosity. 
The Froude number $\text{Fr}=a_K/g$ quantifies the relevance of gravity, where $a_K=\eta_K/\tau_K^2$ is the Kolmogorov acceleration and $\eta_K = \left( \nu^3 / \langle \epsilon \rangle \right)^\frac{1}{4}$ the Kolmogorov length.
The last parameter is the number density relative to the Kolmogorov length $\rho_\text{N}=\frac{N}{\eta_K^3}$. The Kolmogorov scales are determined by the mean dissipation rate. For our analysis, we also define the local Kolmogorov scales based on the volume-averaged dissipation rate, i.e., $\tau_{K,r}=\sqrt{\nu/\epsilon_r}$, $\eta_{K, r}=\left(\nu^3/\epsilon_r\right)^{1/4}$, and $a_{K,r}=\eta_{K,r}/\tau_{K,r}^2$. Therefore, fluctuations in $\bar\epsilon_r$ directly affect the local Stokes number $\text{St}_r=\tau_d/\tau_{K,r}$ and Froude number $\text{Fr}_r=a_{K,r}/g$ on the scale of cloud parcels, see Fig.~\ref{fig:upper_per}.

As collisions are a random process, collisional growth calls for a statistical description.
The following general master equation describes  the time-dependent probability to find droplets of size $n$

\begin{align}
\dot P_n(t) &= j^\text{in}_n(t) - j^\text{out}_n(t)\label{eq:inout},    
\end{align}
where $j^\text{in}_n(t)$ and $j^\text{out}_n(t)$ are influxes and outfluxes, respectively, to be specified.
In the Markovian case of memoryless collisions, one can describe these fluxes by constant collision kernels between $n$ and $n^\prime$ within the established Smoluchowski equation~\cite{Pumir2016, Grabowski2013}. 
For our analytical computations, we assume that droplets most likely collide with droplets of their initial size, i.e., $n^\prime = 1$. 
As a result, droplet size becomes a discrete variable $n>0$ of multiples of their initial size with incremental increase and a direct relation between influx and outflux, $j_{n+1}^{\text{in}}=j_{n}^{\text{out}}$.
In addition, we assume that droplets grow effectively in a statistically stationary background distribution, i.e., the reservoir of background droplets is not reduced by collisions.
As a result, the Smoluchowski equation is linear in the evolving droplet distribution.
This translates to a simple master equation with constant collision rates $\lambda_{n}$~\cite{Telford}

\begin{align}
    \dot P^\text{M}_n(t) &=\lambda_{n-1}P^\text{M}_{n-1}(t) -  \lambda_{n}P^\text{M}_{n}(t)\,,
    \label{eq:smol}
\end{align}
where the superscript $\text{M}$ denotes the Markovian case.
Using variation of constants one obtains a recursive relation for $P^\text{M}_n$ that can be expressed for our initial condition $P_n(0) = \delta_{n1}$ as (see SI) 

\begin{align}
    P^{\text{M}}_n(t,\boldsymbol{\lambda}) & = 
    \int_{-\infty}^t \!\!\!\!\! dt_n\, \cdots \int_{-\infty}^{t_2}\!\!\!\!\! dt_1\,
    \delta(t_1)\nonumber\\ &\times \prod_{i=1}^{n-1}\left[\lambda_i e^{-\lambda_i (t_{i+1}-t_{i})}\right]e^{-\lambda_n(t-t_{n})} ,
    \label{eq:sol_markov}
\end{align}
where the tuple $\boldsymbol{\lambda}=(\lambda_1, ..., \lambda_n)$ specifies the constant collision rates of droplets up to size $n$.
This can also be solved explicitly to yield~\cite{Telford}

\begin{align}
   P^{\text{M}}_n(t,\boldsymbol{\lambda})&=\sum^{n-1}_{l=1}\left[\frac{\prod_{i=1}^{n-1}\lambda_i}{\underset{\,\,\,\,\,\,i\neq l}{\prod_{i=1}^{n}}(\lambda_{i}-\lambda_l)}\left(e^{-\lambda_lt}-e^{-\lambda_nt}\right)\right]\, .\label{eq:Telford}
\end{align}
This Markovian solution assumes that collision times are uncorrelated, an approximation that is not necessarily true in turbulence, as previously shown in \cite{Bec2016}.

To investigate to what extent this assumption holds in cloud conditions, we turn to DNS.
For our numerical investigations, we use the pseudospectral fluid solver TurTLE~\cite{Lalescu2022}.
We consider different dissipation rates that are expected to occur in cloud conditions.
To increase the statistics of correlated collisions in our simulations, we choose a physical droplet number density $1000\,\text{cm}^{-3}$ at the upper limit of what might occur in clouds (see SI for a discussion of how simulation parameters relate to the physical characteristics of clouds~\cite{Pruppacher2010}). 
In the main text, we focus on the mean volume-averaged dissipation rate $\bar\epsilon_r=1$ ($\text{St}_r=0.1$, $\text{Fr}_r=0.15$, $\rho=0.16N\eta_{K,r}^{-3}$), as well as a large volume-averaged dissipation rate $\bar\epsilon_r=9$ ($\text{St}_r=0.3$, $\text{Fr}_r=0.41$, $\rho_n=0.03N\eta_{K,r}^{-3}$) to characterize the strong fluctuations in the 2.6\% most-dissipative parcels.

\begin{figure}
    \centering
    \includegraphics[width=\columnwidth]{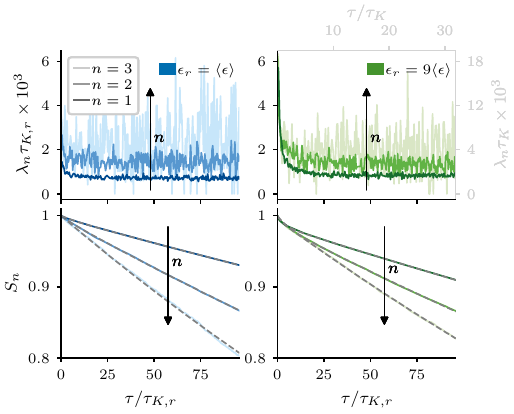}
    \caption{Collisions in turbulence are non-Markovian but can be approximated as a superposition of Poisson processes with different timescales.
    Data from DNS (see SI for parameters) with normalized volume-averaged dissipation rate $\bar\epsilon_r=1$ (blue) and $\bar\epsilon_r=9$ (green).
    (Top) Conditional collision rate for droplets of size $n$. 
    The increase at short times implies temporal correlations and violates the memoryless assumption of Poisson processes with a constant rate.
    (Bottom) The survival probability can be well approximated by a sum of two exponential functions (gray dashed lines). 
    Note that we nondimensionalized with the local Kolmogorov time $\tau_{K,r}$. On the left, this is identical to mean conditions $\tau_K$. 
    On the right, the local and global Kolmogorov times are different and we show the additional axes nondimensionalized by $\tau_K$ in gray.
    }
    \label{fig:self_consistent}
\end{figure}

\section*{The effect of short-time correlations}
To quantify the effect of correlations, we measure the conditional collision rate $\lambda_n(\tau)$ for all droplets of size $n$ for which a collision has occurred at $\tau=0$.
This conditional collision rate is connected to the survival probability (see, e.g., ~\cite{kleinbaum1996survival})

\begin{equation}
  S_n(\tau)=e^{-\int_0^\tau\lambda_n(\tau')d\tau'} \quad\text{for}\quad \tau\geq 0,
  \label{eq:survival}
\end{equation}
which quantifies the fraction of droplets of size $n$ that have not collided again by time $\tau$ (Fig.~\ref{fig:self_consistent}).
If consecutive collisions are not correlated, then $\lambda_n(\tau)=\lambda_n$ is constant, which results in  exponential survival probabilities $S_n(\tau)=e^{-\lambda_n\tau}$.
Since we prepare the system initially with particles of size $n=1$, the first collision cannot be correlated to a previous collision.
This implies that $\lambda_1$ is -- except for small effects of a slowly reducing number of droplets -- constant.
Our analyses further reveal that for the global mean, $\bar\epsilon_r=1$, there are correlations in the collision rates with a slight increase of $\lambda_n(\tau)$ at very short times of $\mathcal{O}(\tau_{K,r})$.
These correlations, though, have no notable effect on $S_n(\tau)$ (Fig.~\ref{fig:self_consistent}, left).
By dimensional arguments, one may expect that in terms of the local Kolmogorov units the results are similar for higher volume-averaged dissipation rates.
Instead, we find for large dissipation, here $\bar\epsilon_r=9$, a stronger time dependence in $\lambda_n(\tau)$ and strong deviations from the exponential shape of $S_n(\tau)$. 
This points to the increased relevance of inertial effects such as clustering, (Fig.~\ref{fig:upper_per}) and implies impactful correlations between collision times of $\mathcal{O}(10\tau_{K,r})$ (Fig.~\ref{fig:self_consistent}, right). 
We can capture the correlations by approximating the survival probability by a superposition of two exponentials, i.e., the superposition of two Poisson processes with different (constant) collision rates,

\begin{align}
    S_n(\tau)=A_n^\mathrm{slow}e^{-\lambda_n^\mathrm{slow}\tau} + A_n^\mathrm{fast}e^{-\lambda_n^\mathrm{fast}\tau}\,,
    \label{eq:example_superpos}
\end{align}
where $A_n^\mathrm{slow} + A_n^\mathrm{fast} = 1$ (see SI for more details on fitting parameters).
We thus find evidence of correlations that violate the assumptions underlying \eqref{eq:smol}. 
What remains to be quantified is whether they are relevant to droplet growth.

To investigate the relevance of correlations, we use a non-Markovian stochastic framework similar to the one established in \cite{Bec2016} that allows us to include correlations in our statistical description explicitly\footnote{Note that in our approach, we keep the full memory of all past collisions by means of a nested integration different from a single convolution with an effective memory kernel used in \cite{Bec2016}.}. 
Due to the time-dependent collision rates, we cannot start from the Markovian master equation, \eqref{eq:smol}, but need to derive a new master equation and corresponding solution for droplet growth from monodisperse initial conditions with non-Markovian influx and outflux.

Let us begin with the special case of droplets of size $n=1$.
Since we prepare the system with a monodisperse droplet distribution at time $t_0=0$, i.e., $P_n(0)=\delta_{n,1}$ and $P_1(0)=1$, the probability of finding droplets of size $n=1$ is directly given by the survival probability $P_1(t)=S_1(t-t_0)$.
For $n=1$, an influx occurs only during preparation as an initial condition, i.e., $j_1^\text{in}(t)=\delta(t-t_0)$, such that all later changes of probability $\dot{P}_1(t)$ in \eqref{eq:inout} are created by the outflux, i.e., $j_1^\text{out}(t)=-\dot{S}_1(t-t_0)$.
From \eqref{eq:survival} follows that $j_1^\text{out}(t)= S_1(t-t_0)\lambda_1(t-t_0)$, which reveals that the outflux is determined by the (time-dependent) collision rate of those droplets that entered state $n=1$ at time $t_0$ and survived until time $t$.

We can generalize this intuition to the case $n>1$.
We recall that the influx is determined by the outflux of one size below $j_n^\text{in}(t) = j_{n-1}^\text{out}(t)$.
As we noticed for $n=1$, the outflux is determined by the collision rate of droplets that entered at time $t_{n}$ and survived until time $t$, which corresponds to $S_n(t-t_n)\lambda_n(t-t_n)=-\dot{S}_n(t-t_{n})$.
This needs to be weighted with all past influxes to yield

\begin{align}  j_n^{\text{out}}(t) = -\int_{-\infty}^t j_{n}^\text{in}(t_{n})\dot{S}_{n}(t-t_{n})dt_{n}.\label{eq:iterative_fluxes}
\end{align}
Notice that for $n=1$, we obtain an integral over influx times $t_1$ that reduces to our above expression due to $j_i^\text{in}(t_1)=\delta(t_1-t_0)$.
For $t_0=0$, we can insert the initial condition into \eqref{eq:iterative_fluxes}, and by using  
$j_n^\text{in}(t) = j_{n-1}^\text{out}(t)$ we iteratively find

\begin{align} j_{n}^{\text{in}}(t_{n})=&\, 
\int_{-\infty}^{t_{n}} \!\!\!\!\! dt_{n-1}\, \cdots \int_{-\infty}^{t_2}\!\!\!\!\! dt_1\, \delta(t_1)
\left[\prod_{i=1}^{n-1} -\dot{S}_{i}(t_
    {i+1}-t_{i})\right].\label{eq:iterative_fluxes_delta_initial} 
\end{align}

Using \eqref{eq:iterative_fluxes} in the master equation, \eqref{eq:inout}, it generalizes for non-Markovian collision rates to the form
\begin{align}
    \dot{ P}_n(t) &=j_n^\text{in}(t) +\int_{-\infty}^t\,j^\text{in}_{n}(t_{n})\dot{S}_n(t-t_{n})dt_{n}\,.    \label{eq:evolv_with_corr}
\end{align}
This is solved by 

\begin{align}
P_n(t)&=\int_{-\infty}^t\,j^\text{in}_{n}(t_{n})S_n(t-t_{n})dt_{n}\,,
\label{eq:sol_no_super}
\end{align}
as one can verify by inserting back into \eqref{eq:evolv_with_corr} and applying the Leibniz integral rule to evaluate the time derivative on the left-hand side.
Moreover, this solution connects to our intuition that the probability of finding droplets in state $n$ is determined by integrating over the past influx from state $n-1$ that survived until time $t$.

Also, note that the evolution equation \eqref{eq:sol_no_super} reduces to \eqref{eq:smol} for constant collision rates. To see that we take an exemplary look at the outflux $j_n^\text{out}$, see \eqref{eq:iterative_fluxes}, 
where we can perform the derivative $\dot{S}_{n}$ to pull out the now constant collision rate

\begin{align}
j_n^\text{out}(t)=&-\int_{-\infty}^t j_{n}^\text{in}(t_{n})\dot{S}_{n}(t-t_{n})dt_{n}\\
=&\lambda_{n}\int_{-\infty}^t j_{n}^\text{in}(t_{n})S_{n}(t-t_{n})dt_{n}
=\lambda_{n}P_{n}(t).
\end{align}
In the last step, we identified \eqref{eq:sol_no_super}.
As $j_{n}^\text{in}=j_{n-1}^\text{out}$, this formulation agrees with \eqref{eq:smol}, showcasing the reduction for constant collision rates.

To solve \eqref{eq:sol_no_super} with its recursive convolutions, we generalize our empirical observation that the survival probability can be expressed as a superposition of exponentials, \eqref{eq:example_superpos}.
Taking a continuum of exponentials yields the Laplace transform

\begin{align}\label{eq:surv_superpos}
    S_n(\tau) = \int_0^\infty \hat{S}_n(\hat\lambda_n) e^{-\hat\lambda_n \tau}d\hat\lambda_n\,.
\end{align}
where $\hat{S}_n(\hat\lambda_n)$ corresponds to the weight of a Poisson process with constant collision rate $\hat\lambda_n$.
Using the superposition \eqref{eq:surv_superpos}, one can explicitly solve the convolution integrals (stemming from the iterative pattern of the fluxes \eqref{eq:iterative_fluxes}) such that the solution \eqref{eq:sol_no_super} becomes 
\begin{figure}
    \centering
\includegraphics[width=\columnwidth]{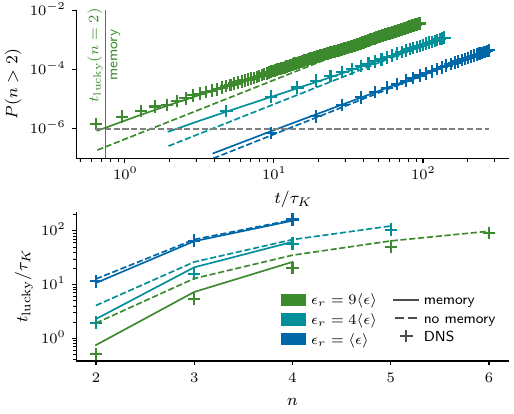}
    \caption{
    (Top) Probability of finding droplets of size $n>2$:
    We compare our non-Markovian stochastic framework using the empirical fit to $S_n$ to turning off correlations by removing the fast-decreasing exponential and our DNS data of droplet growth. 
    This shows that memory effects can further accelerate growth at short times. 
    In the case of $\epsilon_r=9\langle\epsilon\rangle$, the time $t_{\text{lucky}}(n=2)$ where the one-in-a-million fastest droplets surpass a given size $n=2$ decreases by about 50\%. 
    (Bottom) $t_{\text{lucky}}$ for different dissipation rates and as a function of the droplet size: we compare our DNS data to our framework with and without memory effects. 
    The relative effects of memory on $t_{\text{lucky}}$ decrease with the volume-averaged dissipation rate and with increasing droplet size. 
    }
    \label{fig:self_consistent2}
\end{figure}

\begin{align}
P_n(t)&=   
\int_{0}^\infty \!\!\!\! d\lambda_n\, \cdots\int_0^{\infty} \!\!\!\!\! d\lambda_1
\prod_{i=1}^{n}\hat{S}_i(\hat\lambda_i)\nonumber\\
&\times  
\int_{-\infty}^t \!\!\!\!\! dt_n\, \cdots \int_{-\infty}^{t_2}\!\!\!\!\! dt_1\, \delta(t_1)
\nonumber\\ &
\times \prod_{i=1}^{n-1}\left[\hat\lambda_i e^{-\hat\lambda_i (t_{i+1}-t_{i})}\right]e^{-\hat\lambda_n(t-t_{n})}\\
&=
\int_{0}^\infty \!\!\!\! d\lambda_n\, \cdots\int_0^{\infty} \!\!\!\!\! d\lambda_1
\prod_{i=1}^{n}\hat{S}_i(\hat\lambda_i)P_n^M(t,\hat{\boldsymbol{\lambda}}),
\label{eq:solution_superposition}
\end{align}
where we identified the Markovian solution within the integral, cf. \eqref{eq:sol_markov}.
Hence, the solution to \eqref{eq:evolv_with_corr} with memory effects becomes a superposition of solutions without memory effects. This also provides a procedure to turn correlations on and off by choosing the number of Poisson processes in \eqref{eq:surv_superpos}. 
Since collisions in our simulations are well captured by the superposition of two Poisson processes (Fig.~\ref{fig:self_consistent}), we can use \eqref{eq:solution_superposition} by inserting

\begin{align}
    \hat{S}_n(\hat\lambda) = A_n^{\text{slow}}\delta(\hat\lambda-\lambda_n^\text{slow})+A_n^{\text{fast}}\delta(\hat\lambda-\lambda_n^\text{fast})\,.
\end{align}
This leads to a solution with memory effects as a sum of the known solutions $P_n^\text{M}$ for all possible combinations of constant collision rates: 

\begin{align}
P_n(t)=&\left[\prod_{j=1}^n(A_j^\text{fast}\delta_{\hat{\lambda}_j,\lambda_j^{\text{fast}}}+A_j^\text{slow}\delta_{\hat{\lambda}_j,\lambda_j^{\text{slow}}})\right] P_n^M(t,\hat{\boldsymbol{\lambda}}),
\label{eq:solution_superposition_sum}
\end{align}
where we can easily deactivate correlations for any $j$ by setting $A_j^\text{fast}=0$ and $A_j^\text{slow}=1$.

Our non-Markovian stochastic framework thereby enables a quantitative assessment of the effect of correlations on the growth of lucky droplets.
To demonstrate this, we study the evolution of the probability of observing droplets with at least two collisions, i.e., $n>2$  (Fig.~\ref{fig:self_consistent2}, top). 
One can see that the data from the DNS (data points) is well approximated by our solution, \eqref{eq:solution_superposition}, with the empirical superposition of two Poisson processes, \eqref{eq:example_superpos} (solid lines).
When we turn off correlations (dashed lines), i.e., solving \eqref{eq:solution_superposition} only with the dominant Poisson process, we observe deviations at short times that become more pronounced with increasing volume-averaged dissipation.
This implies that correlations can accelerate droplet growth on short timescales but the effect strongly depends on sufficiently high dissipation and number density (see SI).
However, the relative speedup of the $10^{-6}$ fastest growing droplets decreases significantly for larger droplet sizes, see Fig.~\ref{fig:self_consistent2} (bottom). 
Hence, we conclude that correlations can  accelerate growth but can be considered small corrections for the growth of large droplets. This additionally solidifies the assumption that collisions are sufficiently rare to consider them as statistically independent events, see e.g.~\cite{Kostinski2005, wilkinson2023quantifying}.

In the following, we thus focus on the effect of dissipation fluctuations alone, keeping in mind that the effect we find would be even larger when short-time correlations were included.

\begin{figure}
     \centering
     \includegraphics[width=\columnwidth]{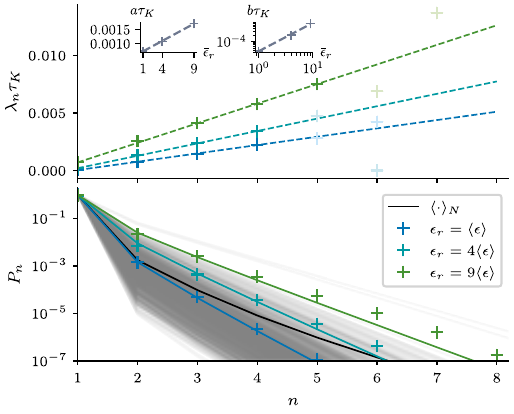}
     \caption{%
     Time-dependent probability distribution of droplet sizes can be approximated by the toy model that integrates \eqref{eq:smol} with an effective $\lambda_n(\bar\epsilon_r)$ parameterized by DNS (data points).
     (Top) Linear approximation of collision rate as a function of collisions, where faded data points were excluded from the fit due to insufficient statistics.
     The inset shows slope $a$ and offset $b$ of our linear approximation that are approximated by a linear function and power-law, respectively.
     (Bottom) Probability distribution at the end time of simulation to benchmark toy model with numerical simulations. 
     The toy model can be easily evaluated for an ensemble drawn from $P(\bar\epsilon_r)$ (gray lines) to find that the ensemble mean (black line) has a higher probability for larger droplets.
     }
     \label{fig:fit}
     \end{figure}

\section*{Toy model to capture dissipation fluctuations}
To estimate the effect of a time-dependent volume-averaged dissipation rate on the time it takes to bridge the size gap, we consider, as a simple toy model, an ensemble of small-scale cloud parcels with linear droplet growth.
Conceptually, this is similar to the eddy-hopping approach, which models supersaturation fluctuations of cloud parcels just that here, we consider dissipation fluctuations instead of supersaturation fluctuations \cite{grabowski2017broadening}.

For each parcel, we model droplet growth by numerically solving a linear master equation, similar to \eqref{eq:smol}, but with a collision rate that depends on time to model the fluctuations in volume-averaged dissipation rates:
\begin{align}
    \dot P_n(t) &= \lambda_{n-1}(t)P_{n-1}(t) -  \lambda_{n}(t)P_{n}(t)\nonumber\\
    \lambda_n(t) &= g(\bar\epsilon_r(t),n)\,.
    \label{eq:toy1}
\end{align}
Here, we model the collision rates as a function $g$ of droplet size and a slowly time-varying local dissipation rate (at fixed scale $r$), which is constrained in the following based on our DNS results. 

In general, the collision rates involve multiple processes \cite{Grabowski2013} such as clustering \cite{Maxey1987, Shaw2003}, sling effect \cite{Falkovich2002, Wilkinson2005}, and differential settling \cite{Ayala2008}. 
We observe, however, an approximately linear dependence of the collision rate on the droplet size $n$, see (Fig.~\ref{fig:fit}), albeit for very small $n$ accessible to our numerical simulations. 
Hence, we make the crude and simplifying ansatz

\begin{align}
    g (\bar\epsilon_r,n)=a(\bar\epsilon_r)(n-1)+b(\bar\epsilon_r)\,,
    \label{eq:coll_rate}
\end{align}
where the parameters depend on $\bar\epsilon_r$ (Fig.~\ref{fig:fit}).

Based on our DNS results, we fit $a$ to depend linearly on the volume-averaged dissipation $a=\left[1.23(2)\times10^{-4}\bar\epsilon_r+6.0(2)\times10^{-4}\right]\tau_K^{-1}$ and $b$ as a power-law $b=3.5\times10^{-5}(4)\bar\epsilon_r^{1.33(8)}\tau_K^{-1}.$ The functional dependence of these parameters is, of course, rather a choice than a fit, given the few data points we have, but we believe it is sufficient that our toy model will capture the essence of how dissipation fluctuations may affect growth statistics.
Indeed, Fig.~\ref{fig:fit} (bottom) shows that matching $\bar\epsilon_r$ with our reference simulations and numerically solving \eqref{eq:smol} (solid colored lines) yields good agreement with our numerical simulations (data points), where deviations in the tails are expected for larger $\bar\epsilon_r$ from neglecting correlations.

This toy model now allows us to rapidly generate droplet size distributions for a full ensemble of dissipation rates. 
Drawing dissipation rates from $P(\bar\epsilon_r)$ expected in clouds (cf. Fig.~\ref{fig:upper_per}), we find that the ensemble mean has an increased probability for larger droplets compared to the mean dissipation (Fig.~\ref{fig:fit} bottom).
Since lucky droplets are statistical outliers, we can already conclude that the log-normal fluctuations in dissipation rate increase the size of lucky droplets.
We are left with the question of whether this can significantly accelerate the growth of droplets of sufficient size to bridge the size gap.

\begin{figure}
     \centering
     \includegraphics[width=\columnwidth]{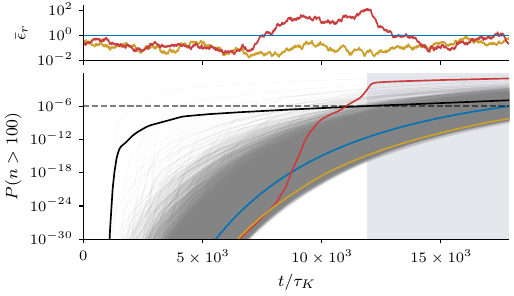}
     \caption{(Top) Two realizations of the volume-averaged dissipation rate drawn from the same distribution - the one with the lowest and highest dissipation rate among $10^5$ randomly generated realizations.
     The corresponding dissipation rate may - as one of the example realizations shows - experience significant spikes.
     (Bottom) The probability that $n>100$, i.e., a droplet had at least 100 collisions as a function of time within our toy model, assuming a more realistic number density of $200\,\text{cm}^{-3}$. 
     The $10^5$ ensemble realizations are shown in gray, whereas we have the ensemble mean in black.
     Notice that large outliers (logarithmic $y$-scale) dominate the mean, e.g., the high-dissipation realization (red) towards the end.
    }
     \label{fig:corr_nocorr}
\end{figure}

Our toy model now provides an affordable approach to extend the evolution of droplet growth to longer times.
On these timescales, however, the volume-averaged dissipation rate will fluctuate notably with an unknown timescale.
A timescale can be estimated by dividing the energy contained in the flow by the dissipation rate. 
On the scale of an individual parcel, this yields a few seconds.
Unfortunately, the precise timescales in clouds remain unknown because measurements of the volume-averaged dissipation rate on the relevant spatial scales ($\sim 1\text{m}$) are currently not reachable within clouds \cite{schroeder2023}.
In the following, we assume a timescale of about $\tau_\epsilon=19\,\text{s}$ (we also tested 10s and 40s with qualitatively similar results, see SI), and construct time-dependent logarithmic dissipation rates as an Ornstein-Uhlenbeck process with corresponding variance and mean from the refined similarity hypothesis to recover the log-normal distribution $P(\bar\epsilon_r)$, \eqref{eq:lognorm}.
We obtain this process by considering the logarithm of the dissipation $\bar\epsilon_r$, $\xi$ (with $\bar\epsilon_r = e^{\xi(t)}$), as a Gaussian process with the following correlation property:
\begin{align}\left\langle(\xi(t)+\sigma^2/2)(\xi(t+\tau)+\sigma^2/2)\right\rangle=\sigma^2e^{-\tau/\tau_\epsilon}\,.
\label{eq:dissipation_time}
\end{align}
 
From the $10^5$ realizations that we generate, two are shown in Fig.~\ref{fig:corr_nocorr} with minimal (yellow) and maximal (red) time-averaged dissipation $\langle\bar\epsilon_r\rangle_t\approx1/5$ and $\langle\bar\epsilon_r\rangle_t\approx7$, respectively, for a time series of about 40 times the timescale of the volume-averaged dissipation rate.

To estimate the effect of dissipation fluctuations on an ensemble of cloud parcels via \eqref{eq:toy1}, we thus make our empirical ansatz \eqref{eq:coll_rate} and model the time evolution of the volume-averaged dissipation $\bar\epsilon_r(t)$ as a stochastic process that is characterized by \eqref{eq:dissipation_time}.
This way we obtain different realizations of parcels and their droplet growth (Fig.~\ref{fig:corr_nocorr}, bottom).
To consider a typical droplet number density of $200\, \text{cm}^{-3}$ instead of the upper bound $1000\,\text{cm}^{-3}$ used in our main simulation, we rescale the collision rates proportionally to the density.
Then, we can use our toy model to infer how much time is needed until lucky droplets bridge the size gap.
Starting with an initial size corresponding to $12.5\,\mu\text{m}$, it is about 100 collisions that suffice to bridge the gap and reach about $50\,\mu\text{m}$.
Considering no fluctuations and only mean dissipation (blue), the fraction of droplets with more than 100 collisions would exceed the lucky-droplet threshold of $10^{-6}$ at about $18\times10^3\tau_K$ (corresponding to 350s, with the assumed cloud parameters)\footnote{Note that this is less than 6 min, which is way below the onset time of rain of 30 min.
Two reasons for such fast times are
i) that we do not account for the time it takes to create droplets of 12.5 $\mu\text{m}$, and
ii) that the collision efficiency for small droplets can be significantly reduced, thereby slowing down the collisional growth.
While these factors could be, in principle, incorporated into our non-Markovian stochastic framework (similar to \cite{wilkinson2023quantifying} for the case of constant collision rates), they would require empirical evidence that still needs to be provided.
However, we expect that the important qualitative results from our toy model will transfer to more realistic conditions.}. 
Considering the ensemble mean (black), representative of a full cloud, we would find the size gap to be bridged already at $12\times10^3\tau_K$ (corresponding with the assumed cloud parameters to 230s), implying that in our model with the chosen parameterization, dissipation fluctuations speed up rain formation by about 33\% (see SI for controls).
The individual realizations provide more insights into how this speed-up occurs:
The ensemble average is dominated by a few statistical outliers.
This is exemplified by looking at the $\bar\epsilon_r(t)$ realization with the highest mean dissipation rate (red), where one can see that, while initially even below mean conditions, the number of large droplets shoots up precisely when the dissipation rate features extreme peaks. 
This implies that rain droplet growth is drastically accelerated in cloud parcels with temporally high dissipation rates and that fluctuations of the volume-averaged dissipation could be crucial to bridge the size gap.

\section*{Summary and Conclusions}

In summary, we established a systematic approach to account for short-time correlations between individual collisions and spatio-temporal dissipation fluctuations. 
We showed that these short-time correlations can accelerate the growth of droplets on short timescales for high particle density and high dissipation, but that this is subleading to the overall acceleration due to fluctuations of the volume-averaged dissipation rate for large droplets.
Using a toy model to capture dissipation fluctuations due to intermittency, we found that strong bursts of the volume-averaged dissipation rate can dominate the formation of fast-growing droplets to bridge the size gap much faster than assuming a constant volume-averaged dissipation rate.
This underscores the potentially crucial contribution of dissipation fluctuations to bridge the size gap sufficiently fast for the onset of rain within 30 min in warm clouds.

\begin{acknowledgments}
This work was supported by the Fraunhofer-Max-Planck Cooperation Program through the TWISTER project and a grant of the European Research Council (ERC) under the European Union’s Horizon 2020 research and innovation programme (Grant agreement No. 101001081). 
J. Z. was supported by the Joachim Herz Stiftung. 
T. B. is grateful for the support through a fellowship of the IMPRS for Physics of Biological and Complex Systems.
We thank Jeremy Bec and Bernhard Mehlig for helpful discussions. 
We thank Cristian C. Lalescu and B\'erenger Bramas for their support and development of the TurTLE code used in this study. 
Computational resources from the Max Planck Computing and Data Facility and the Max Planck Institute for Dynamics and Self-Organization  as well as the support by the Max Planck Society are gratefully acknowledged.
\end{acknowledgments}

\bibliography{reference}

\clearpage
\newpage
\onecolumngrid

\begin{center}
\textbf{\large Supplementary Information}
\end{center}
\setcounter{equation}{0}
\setcounter{figure}{0}
\setcounter{table}{0}
\setcounter{page}{1}
\makeatletter
\renewcommand{\theequation}{S\arabic{equation}}
\renewcommand{\thefigure}{S\arabic{figure}}
\renewcommand{\thetable}{S\arabic{table}}

\section{Solution of evolution equation with constant collision rates}

This section details how we solve the master equation, Eq.~(5) in the main text. 
For completeness, we consider the master equation

\begin{align}
    \dot P^{\text{M}}_n(t) &= \lambda_{n-1}P^{\text{M}}_{n-1}(t) -  \lambda_{n}P^{\text{M}}_{n}(t)\,,
    \label{eq:sup_smol}
\end{align}
which describes the growth of droplets that collide with a constant rate in a monodisperse and statistically stationary background distribution.
This can be solved by variation of constants: Here $\dot{P}_n^M(t) + \lambda_{n}P^{\text{M}}_{n}(t) = 0$ is the homogeneous differential equation that is solved by $P^{M,\text{hom}}_n(t)=c e^{-\lambda_n t}$. 
The method of the variation of constant with this homogeneous solution and the inhomogeneity $\lambda_{n-1}P^{\text{M}}_{n-1}(t)$ then results in the following expression for the full solution

\begin{align}
    P^\text{M}_n(t) &= e^{-\lambda_n t}\left(c_n + \int_{0}^t \lambda_{n-1} P^\text{M}_{n-1}(t_n)e^{\lambda_n t_{n}} dt_n\right) = c_n e^{-\lambda_n t} + \int_0^t \lambda_{n-1}e^{-\lambda_n(t-t_n)}P^\text{M}_{n-1}(t_n) \, dt_n\,.\label{eq:variation_constant}
\end{align}
The integration constant $c_n$ is constrained by the initial condition $P^\text{M}_n(0)=\delta_{n1}$ such that $c_1=1$ and $c_n=0$ for $n>1$.

Let us build up the solution according to this iterative pattern starting with $n=1$. 
Here, the second term in \eqref{eq:variation_constant} vanishes as there is no smaller size than $n=1$. 
This yields, as one might expect, a simple exponential decay for $t\geq 0$. 
In preparation for our non-Markovian treatment, we can also write the Markovian solution with an integral extending over the full past and introduce a $\delta$-influx to model the initial condition

\begin{align}
    P^\text{M}_1(t)= e^{-\lambda_1 t}=\int_{-\infty}^{t}\delta(t_1)e^{-\lambda_1 (t-t_1)}dt_1 \,.
\end{align}
Following the above iterative pattern we obtain for $n=2$ (recall $c_2=0$)

\begin{align}
    P^\text{M}_2(t)= \int_{-\infty}^t\int_{-\infty}^{t_2}\delta(t_1)\lambda_1 e^{-\lambda_2 (t-t_2)}e^{-\lambda_1 (t_2-t_1)}dt_1dt_2\,,
\end{align}
and $n=3$

\begin{align}
    P^\text{M}_3(t)= \int_{-\infty}^t\int_{-\infty}^{t_3}\int_{-\infty}^{t_2}  \delta(t_1)\lambda_2 \lambda_1 e^{-\lambda_3 (t-t_3)}e^{-\lambda_2 (t_3-t_2)}e^{-\lambda_1 (t_2-t_1)}dt_1dt_2dt_3\,.
\end{align}
Following this pattern further, the solution becomes

\begin{align}
    P^\text{M}_n(t)= \int_{-\infty}^t...\int_{-\infty}^{t_2}\delta(t_1) \left[\prod_{i=1}^{n-1}\lambda_i e^{-\lambda_i (t_{i+1}-t_i)}\right]e^{-\lambda_n (t-t_n)}dt_1 ... dt_n,
\end{align}
which is provided in the main text.

\newpage
\section{Parameters and simulation setup}
\label{sec:para}

In the foreseeable future, simulations cannot resolve the entire range of scales and complexity of cloud turbulence, where the integral scale is on the order of $\sim100\text{m}$ while the Kolmogorov scale, depending on the dissipation rate, is on the order of 1mm.

Our approach to this challenge is to focus on smaller parcels of cloud turbulence that can be resolved within direct numerical simulations. 
Here, we use the pseudospectral fluid solver TurTLE \cite{Lalescu2022} to create a statistically stationary flow of homogeneous isotropic turbulence in a periodic box. In that flow, we evolve the droplets according to Eq.~(3) in the main text.
Once the droplet distribution reaches a statistically stationary state, we allow droplets to merge upon collisions, where we assume a collision efficiency of one, see section \ref{sec:collision_procedure}.
In total, our simulation box measures $\sim 860$ Kolmogorov lengths in each direction (resolved on a $1024^3$ grid). 
As the box covers a fixed number of Kolmogorov lengths, it may correspond to different physical dimensions depending on the assumed {volume-averaged} dissipation rate within the box.  
Due to intermittency, {this volume-averaged} dissipation rate fluctuates, which can be modeled by the refined similarity hypothesis, see Eq.~(1) in the main text. 
As those fluctuations change the physical dimensions of the simulation box, we also have to vary the number of droplets per box accordingly to preserve the density, see Tab~\ref{tab:par}. 
Additionally, the fluctuations also influence the non-dimensional characteristics that govern the role of inertia and gravity on the dynamics of the droplets, Stokes number and Froude number, respectively, as sketched in Fig.~1 in the main text.    

For our simulations, we choose a mean dissipation rate and a number concentration of droplets at the upper end of the parameter range that is typically observable within clouds \cite{Pruppacher2010}, see Tab.~\ref{tab:typ_values}. 
We initialize the droplets as a monodisperse population, corresponding to a size of $12.5\,\mu\text{m}$, which is in the range of typical droplet sizes.
To capture typical fluctuations of the dissipation rate on the scales of the simulation box, we once take the {volume-averaged} dissipation rate of the simulation corresponding to the global mean as well as four and nine times higher, see Tab.~\ref{tab:par}.
As discussed above, the effective volume of our box changes accordingly. 
So while for $\epsilon_r=9\langle\epsilon\rangle\approx0.36 \text{m}^2\text{s}^{-3}$ the simulation box has a length corresponding approximately to a quarter meter, for $\epsilon_r=\langle\epsilon\rangle\approx0.04 \text{m}^2\text{s}^{-3}$ the box length corresponds to almost half a meter. 
According to the effective volume, the number of droplets in the simulation box $N_\text{sim}$ also varies.

\begin{table}[h!]
    \centering
   \begin{tabular}{c|c|c|c|c}
     mean dissipation rate $\langle\epsilon\rangle\,[\text{m}^2\text{s}^{-3}]$ & viscosity $\nu\,[\text{m}^2\text{s}^{-1}]$& droplet radius $a\,[\mu\text{m}]$ & number density $\rho_{\text{N}}\,[N/\text{cm}^3]$ & density ratio water/air $\rho_L/\rho_0$ \\\hline
     $0.001-0.04$ & $\approx1.5\times10^{-5}$&$<20$& $100-1000$ &$\approx800$
\end{tabular}
    \caption{Typical literature values which govern droplet and turbulence properties \cite{Pruppacher2010}: mean dissipation $\epsilon$, viscosity $\nu$, droplet radius $a$, number density $n$ and density ratio between water and air $\rho_L/\rho_0$}
    \label{tab:typ_values}
\end{table}

\begin{table}[h]
    \centering
    \begin{tabular}{c|c|c|c||c|c|c|c|c|}
         \#&$\epsilon_r\,\left[\frac{\text{m}^2}{\text{s}^3}\right]$&$a\,[\mu\text{m}]$&$\rho_\text{N}\,\left[\frac{N}{\text{cm}^3}\right]$&$\text{St}_r$&$\text{Fr}_r$&$\rho_\text{N}\left[\frac{N}{\eta^3}\right]$&$N_\text{sim}$  \\\hline
         $1$&$0.04$&$12.5$&$1000$&$0.1$&$0.15$&$1.6\times10^{-1}$& $115\times10^6$\\
         $2$&$0.16$&$12.5$&$1000$&$0.2$&$0.41$&$5.5\times10^{-2}$&$40\times10^6$\\
         $3$&$0.36$&$12.5$&$1000$&$0.3$&$0.76$&$3.0\times10^{-2}$&$22\times10^6$
    \end{tabular}
    \caption{Assumed physical parameters such as local dissipation rate, droplet size, and number density in terms of the microscale and the total number of droplets in the simulation domain, and corresponding non-dimensional characteristics. The characteristics of the flow remain unchanged in code units; all runs are conducted on a $1024^3$ grid with a spatial resolution of $k_{\text{max}}\eta\approx3$ and a Reynolds number of $\text{Re}_\lambda\approx200$.}
    \label{tab:par}
\end{table}

\newpage
\begin{figure}[h]
    \centering
    \includegraphics[width=0.4\textwidth]{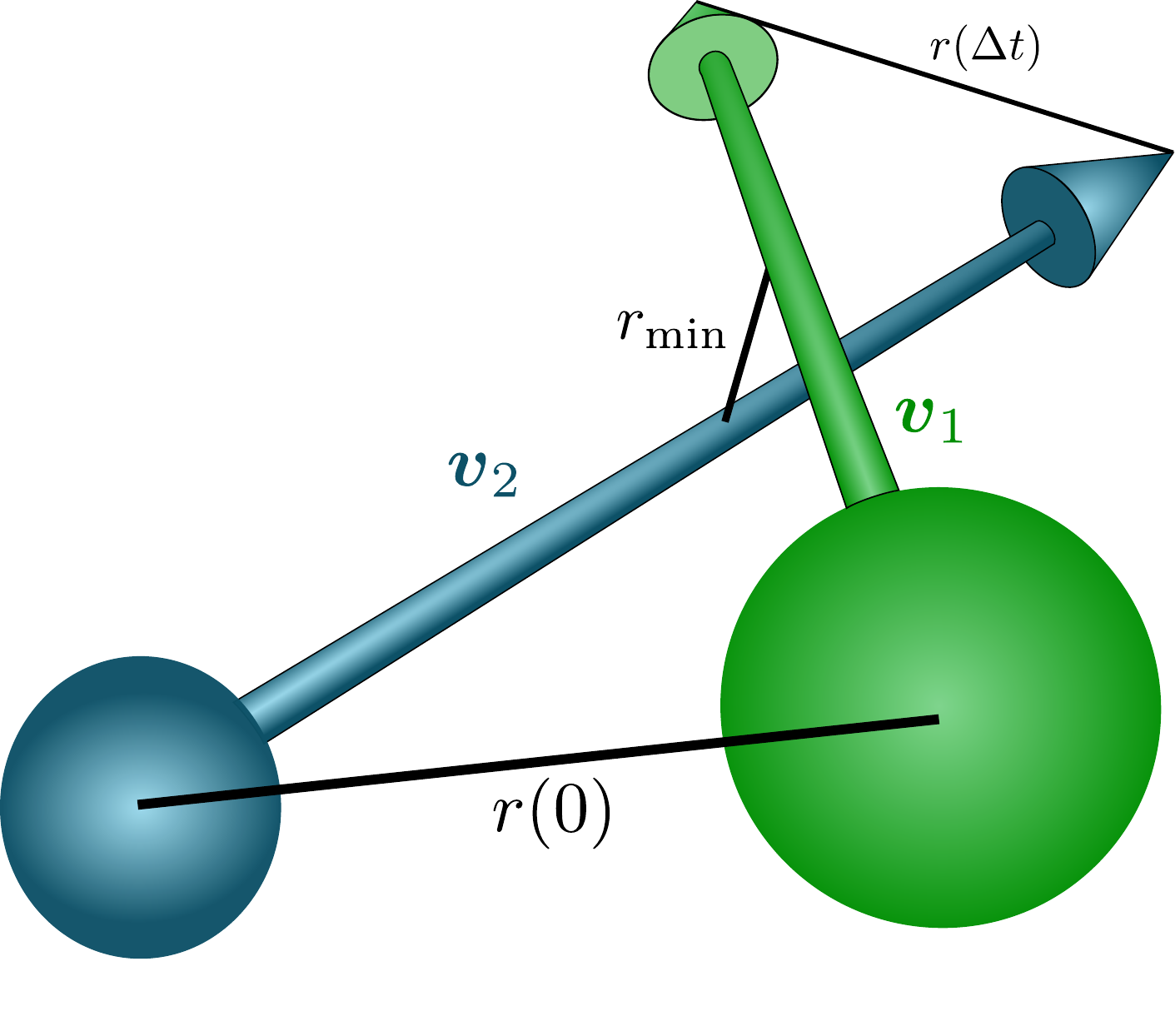}
    \caption{Two particles with velocities $\boldsymbol{v}_{1/2}$ and separation $r(0)$ at the beginning of the timestep. Their separation changes then -- according to the linear extrapolation -- to $r(\Delta t)$ within one timestep, whereas it may reach the minimal distance, $r_{\text{min}}$ in between.}
    \label{fig:my_label}
\end{figure}
\section{Collision procedure}
\label{sec:collision_procedure}

For the detection of collisions of larger particles, it could be simply sufficient to check whether particles overlap at the corresponding timesteps. For an example that uses this approach in our code TurTLE, we refer to \cite{arguedas2022elongation}. 
In that case, the procedure is justified as particles may not move relative to each other more than the collision radius within one timestep. 
However, the droplets we consider are small, and through the described procedure one might miss collisions in between timesteps. 
Therefore, we linearly extrapolate the trajectories every timestep to check for particles coming closer than the collision radius between timesteps (as also done, e.g., in \cite{bhowmick2019direct}). 
We describe the procedure in the following:
Numerically detecting collisions requires first identifying droplets that may collide within one timestep. 
The fluid solver TurTLE \cite{Lalescu2022} features functionality that efficiently provides all pairs of particles below a certain distance, in the following referred to as cutoff distance.
Here, we take the grid spacing as the cutoff distance for our particles, which are significantly smaller than that. 
Given a sufficiently small timestep, they do not travel distances larger than the grid spacing within a single timestep.
Having identified the collision candidates, we can extrapolate their trajectories to check for collisions between the current and next timestep. 
Here, linear extrapolation is sufficient as the small numerical errors do not accumulate (This is in contrast to the time evolution itself, where small errors grow exponentially).

Let two collision candidates have the radii $r_{p,1}$ and $r_{p,2}$, positions $\boldsymbol{x}_1$ and $\boldsymbol{x}_2$, and velocities $\boldsymbol{v}_1$ and $\boldsymbol{v}_2$ at the time  $t=0$. Via extrapolation, we can check for collisions occurring during the following timestep $\Delta t$.
For this, we need to find the minimal distance between the particle centers by considering

\begin{align}
    r_{\text{min}}^2=\min_{\Delta t\ge t\ge 0}[r(t)^2]=\min_{\Delta t\ge t\ge 0}\left[|(\boldsymbol{x}_1+\boldsymbol{v}_1t)-(\boldsymbol{x}_2+\boldsymbol{v}_2t)|^2\right]
\end{align}
Hence, the following condition determines the time when the droplets reach their minimum distance:

\begin{align}
    \frac{d}{dt}\left[ (\boldsymbol{x}_1+\boldsymbol{v}_1t)-(\boldsymbol{x}_2+\boldsymbol{v}_2t)\right]^2=0 \quad 
    \Rightarrow \quad \boldsymbol{r}\cdot\Delta\boldsymbol{v}+(\Delta\boldsymbol{v})^2t=0
\end{align}
which implies

\begin{align}
 t_{\text{min}}=-\frac{\boldsymbol{r}\cdot\Delta\boldsymbol{v}}{(\Delta\boldsymbol{v})^2}\label{eq:t_min} \,.
\end{align}
Here, $\boldsymbol{r}=\boldsymbol{x}_1-\boldsymbol{x}_2$ is the relative distance of the particles and $\Delta \boldsymbol{v}=\boldsymbol{v}_1-\boldsymbol{v}_2$ their relative velocity. Depending on the time $t_{\text{min}}$, when the droplets reach the minimal distance, we can exclude or conclude that a collision will happen within the next timestep.
For $\Delta t>t_{\text{min}}>0$ we have a collision, if $r_{\text{min}}=r(t_{\text{min}})$ fulfills the collision condition, 

\begin{align}
     r_\text{min} < r_{p,1}+r_{p,2} \,.\label{eq:collision_condition}
\end{align}
Using the expression for $t_{\text{min}}$, \eqref{eq:t_min}, we obtain:

\begin{align}
    r_{\text{min}}&=|\boldsymbol{r}(0)+\Delta\boldsymbol{v}t_{\text{min}}|\\&=\sqrt{\boldsymbol{r}^2-\frac{(\boldsymbol{r}\cdot\Delta\boldsymbol{v})^2}{(\Delta\boldsymbol{v})^2}}\,.
\end{align}
For $t_{\text{min}}>\Delta t$, we can also have a collision. While they reach their minimal distance only at a later point, their distance may become smaller than the collision radius during the timestep. 
To confirm this case, we check if at the end of the timestep, the distance $r(\Delta t)$ fulfills the collision condition:

\begin{align}
     r(\Delta t) < r_{p,1}+r_{p,2} \,.\label{eq:collision_condition}
\end{align}

\begin{figure}[h!]
    \centering   \includegraphics[width=0.6\textwidth]{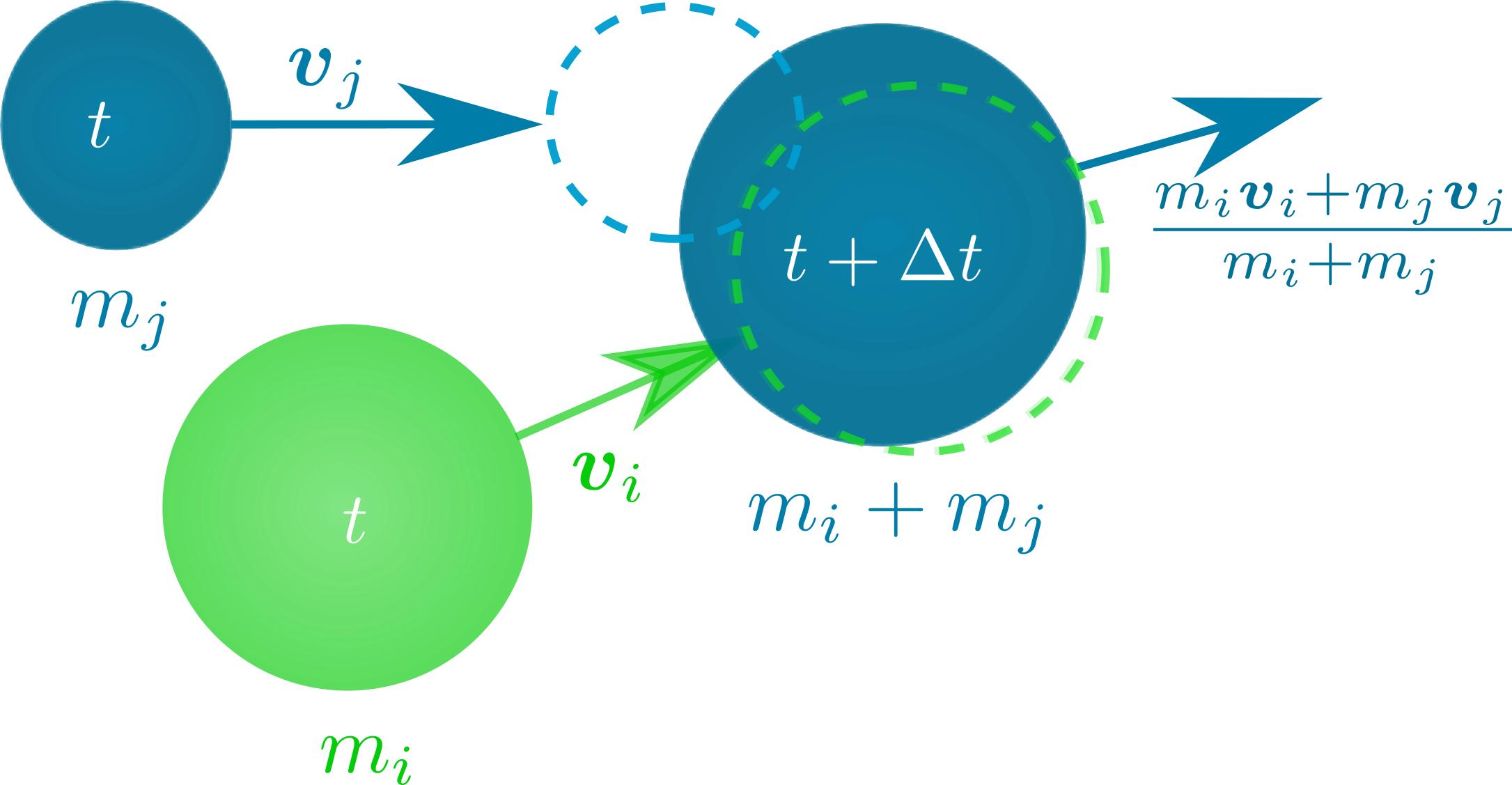}   \caption{ Upon collision of particle $j$ and $i$ (with $i>j$) within one timestep $\Delta t$, they merge where mass and momentum are conserved. According to the collision procedure, the droplet with the smaller index persists while we delete the other one.}
    \label{fig:collsion_procedure}
\end{figure}

If one of the two cases above is fulfilled we conclude that during the timestep a collision occurs. Once we detect a collision, we model the process of the collision itself. 
Here, we change the mass and velocity of the colliding particles according to momentum and mass conservation upon a collision, see Fig.~\ref{fig:collsion_procedure}, where we choose the droplet with the smaller index to grow:
 The position changes to the center of mass of the two colliding particles. This can be summarized as:
 
\begin{align}
m_j\rightarrow m_i+m_j\quad
    \boldsymbol{v}_j\rightarrow \frac{m_i\boldsymbol{v}_i+m_j\boldsymbol{v}_j}{m_i+m_j}\quad
    \boldsymbol{x}_j\rightarrow \frac{m_i\boldsymbol{x}_i+m_j\boldsymbol{x}_j}{m_i+m_j}\quad \text{with}\quad j<i\,.
\end{align}
We add the droplet with the larger index to a list of particles, which we delete before the next timestep. A consistent choice of which particle we delete simplifies the bookkeeping, while it does not make a physical difference.

\newpage
\section{Survival functions for larger droplet sizes}
In Fig.~2 of the main text, we show the extracted survival functions and a fit based on the superposition of two Poissonian processes. 
Consequently, we approximate the different survival probabilities as a sum of two exponentials:

\begin{align}
        S_n(\tau)=A_n^\text{slow}e^{-\lambda_n^\text{slow}\tau}+A_n^\text{fast}e^{-\lambda_n^\text{fast}\tau}\,.
\end{align}
Here, $\lambda_n^\text{slow}$ denotes the collision rate capturing the exponential tail of the survival probability and the long-time behavior while $\lambda_i^\text{fast}$ corresponds to the rate of the additional short-time process. 
$A_n^\text{slow}$ and $A_i^\text{fast}$ quantify the respective weights of the two processes. 
The fit values up to n=4 are shown in Tab.~\ref{tab:fit}, and form the basis of the solutions $P_3$ and $P_4$ in Fig.~3 of the main text according to Eq.~(19) in the main text.

\begin{table}[h]
    \centering
    \begin{tabular}{c||c|c|c}
         &$\epsilon_r=\langle\epsilon\rangle$&$\epsilon_r=4\langle\epsilon\rangle$&$\epsilon_r=9\langle\epsilon\rangle$\\\hline
         $\lambda_1^\text{slow}\,[\tau_K^{-1}]$ &$3.616(8)\times10^{-5}$&$1.999(4)\times10^{-4}$&$7.00(1)\times10^{-4}$\\\hline
         $A_2^\text{fast}$&$0.298(2)$\%&$0.718(2)$\%&  $1.322(3)$\%\\
         $\lambda_2^\text{slow}\,[\tau_K^{-1}]$ & $7.211(2)\times10^{-4}$&$1.3552(6)\times10^{-3}$&$2.548(2)\times10^{-3}$\\
         $\lambda_2^\text{fast}\,[\tau_K^{-1}]$&$1.4(2)\times10^{-1}$&$5.39(5)$&$8.05(9)$\\\hline
         $A_3^\text{fast}$&$0.298(4)$\%&$0.78(1)$\%&$1.294(3)$\%\\
         $\lambda_3^\text{slow}\,[\tau_K^{-1}]$ &$1.459(5)\times10^{-3}$&$2.356(3)\times10^{-3}$&$4.117(4)\times10^{-3}$\\
         $\lambda_3^\text{fast}\,[\tau_K^{-1}]$&$2.1(1)$&$2.17(8)$&$5.02(7)$\\\hline
         $A_4^\text{fast}$&$0.000(4)$\%&$0.36(1)$\%&$0.527(6)$\%\\
         $\lambda_4^\text{slow}\,[\tau_K^{-1}]$ &$2.238(3)\times10^{-3}$&$3.401(4)\times10^{-3}$&$5.761(3)\times10^{-3}$\\
         $\lambda_4^\text{fast}\,[\tau_K^{-1}]$& -&$2.1(5)$& $3.7(3)$
   
    \end{tabular}
    \caption{Fit parameters for the survival functions in Fig.~\ref{fig:fit} for the three different parameter sets. $\lambda_1^{\text{slow}}$ corresponds to the exponential decay rate of the first size. Here we only have one exponential as the first collision cannot correlate with a previous one. Otherwise, we fitted a superposition of a short and long-time process. $\lambda_i^{\text{slow}}$ corresponds to the long-time collision rate while $\lambda_i^{\text{fast}}$ corresponds to the rate of the additional short-time process. Here, $A_i^{\text{fast}}= 1-A_i^{\text{slow}}$ denotes the weight of the short-time process and quantifies the fraction of droplets colliding due to correlations.}
    \label{tab:fit}
\end{table}

\begin{figure}[h]
    \centering
    \includegraphics[width=0.8\textwidth]{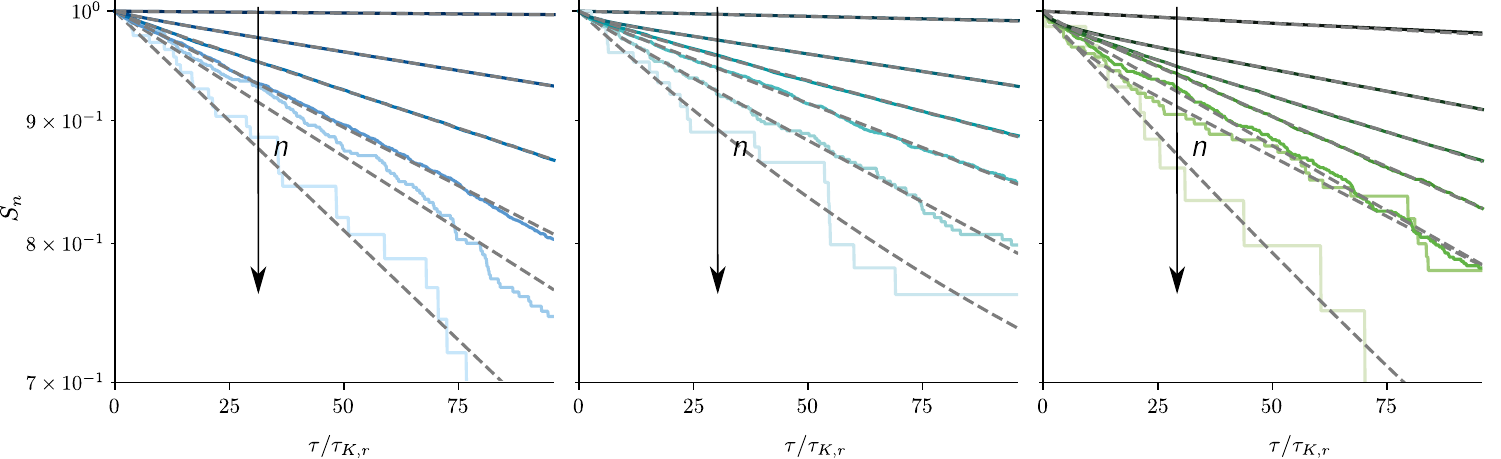}
    \caption{ Survival probabilities for the different local mean energy dissipation rates $\bar\epsilon_r= 1,4,9$ from left to right. With increasing droplet size $n$, the lines become lighter and statistically less converged. In Fig.~4 of the main text, we extrapolate the values of the collision rates $\lambda_n^{\text{slow}}$ associated with the exponential tail.}
    \label{fig:sup_fit}
\end{figure}

In the main text (Fig.~4), we also compare the solution for the droplet population without memory effects, i.e., the solution to the Markovian master equation, $
    \dot P_n(t) = \lambda_{n-1}P_{n-1}(t) -  \lambda_{n}P_{n}(t)\,,
$
constrained by collision rates from the DNS data for a range of droplet sizes.
Accordingly, we need the constant collision rates neglecting the effect of correlations, i.e., the rate corresponding to the exponential tail of the survival function, also for higher values of $n$. For higher values of $n$, we only have a few droplets reaching those sizes and an insufficient statistical basis to fit those values. Figure~\ref{fig:sup_fit} also includes the less converged survival probabilities for higher $n$ complementary to Fig.~2 in the main text. 
In Fig.~4 in the main text, we consider the fit for $\lambda_n^{\text{slow}}$ up to a value of $n$ where individual collisions lead to fluctuations featured as pronounced steps and the fit becomes unreliable. For higher values of $n$, we therefore extrapolate. 

\newpage
\section{Memory effects in DNS with lower droplet number density}
\label{sec:lucky400}
\begin{figure}[h]
     \centering
         \includegraphics[width=0.7\textwidth]{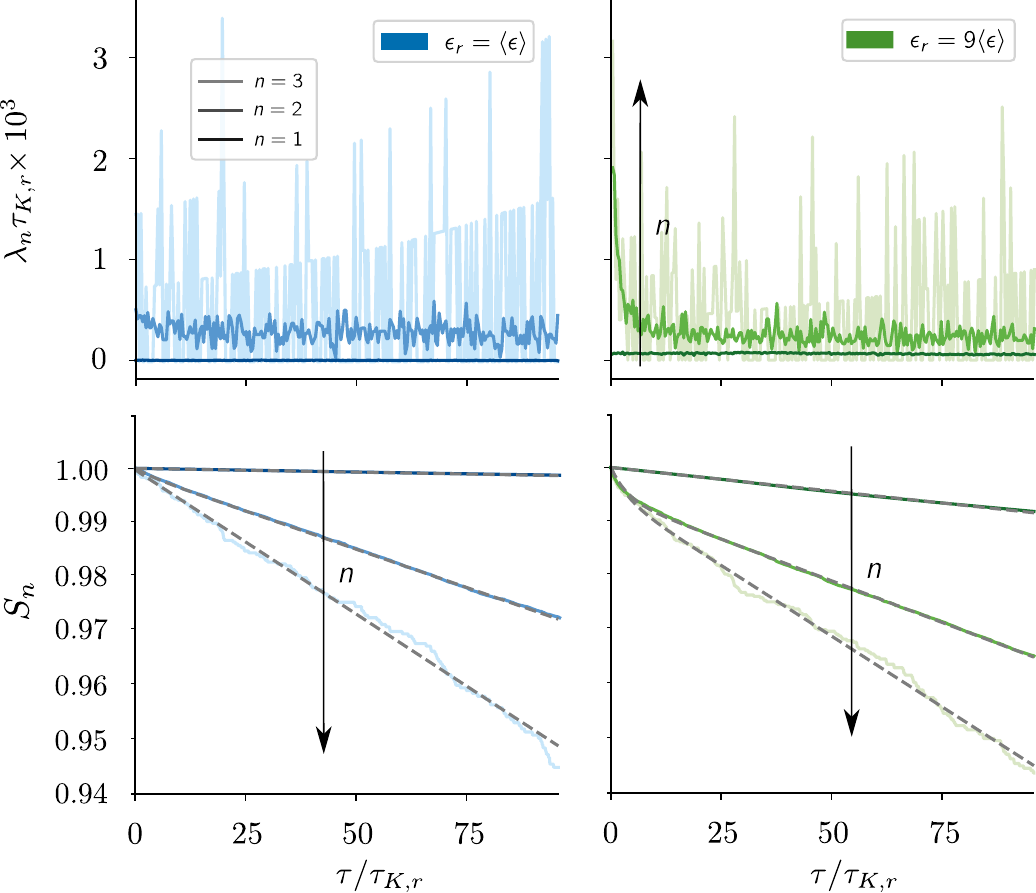}
     \caption{ Collision rates $\lambda_n$ and survival probability $S_n$ for different sizes $n$ for a number density of $400\,\text{cm}^{-3}$: here, $\tau$ denotes the time since the last collision except for $n=1$, where it is the time since the simulation started.
    (top) Except for $\lambda_1$, which is close to constant, all collision rates are, for short times, significantly increased before they reach approximately a constant (up to fluctuations). The relative increase on short times is more pronounced for $\epsilon_r=9\langle\epsilon\rangle$ than for $\epsilon_r=\langle\epsilon\rangle$, while generally, collision rates are higher for a higher local mean energy dissipation rate. 
    (bottom) We fit each survival function by superposing two exponentials, one for the tail and one to capture short-time correlations. For the fit parameters, see Tab.~\ref{tab:fit_lucky_400}. (bottom left) For the local mean $\epsilon_r=\langle\epsilon\rangle$, the survival probabilities all have a shape close to an exponential one. The fast Poisson process has only a weight between 0-0.569\%. (bottom right) For $\epsilon_r=9\langle\epsilon\rangle$, one can visually see the effect of correlations for short times where the fast Poisson process weights up to $0.569\%$.}
     \label{fig:rate_400}
 \end{figure}

Let us consider the  memory effects in collisional droplet growth for a lower droplet number density of $400 \, \text{cm}^{-3}$.
We start with the collision rates and survival functions as shown in Fig.~\ref{fig:rate_400}. Generally, with a lower number of droplets, collisions are significantly less likely and, therefore, statistical convergence is weaker as one can also see by the larger scatter compared to Fig.~2 in the main text. As a consequence, we only show the collision rates and survival functions of the first three sizes. The collision rates still have a clear signature of correlations between consecutive collisions, more pronounced in the case of $\epsilon_r=9\langle\epsilon\rangle$. As for the case of higher number density, the survival functions in the case $\epsilon_r=\langle\epsilon\rangle$ are almost exponential, i.e., memory effects are negligible. In contrast, the survival probabilities for $\epsilon_r=9\langle\epsilon\rangle$ deviate at short times from an exponential behavior through a more rapid decrease. 
The fit of the superposition of two exponential functions -- one capturing the tail and one the memory effects on short times -- shows that the weight of the short-time process $A_i^\text{fast}\sim0.52-0.57\%$ is still notable. 
However, its relevance compared to the case studied in the main text is further reduced. For the fit parameters, we refer to Tab.~\ref{tab:fit_lucky_400}.

 \begin{table}[h]
    \centering
    \begin{tabular}{c||c|c|c}
         \#&$\epsilon_r=\langle\epsilon\rangle$&$\epsilon_r=4\langle\epsilon\rangle$&$\epsilon_r=9\langle\epsilon\rangle$\\\hline
         $\lambda_1^\text{slow}\,[\tau_K^{-1}]$ &$1.43(5)\times10^{-5}$&$8.02(2)\times10^{-5}$&$2.699(2)\times10^{-4}$\\\hline
         $A_2^\text{fast}$&$0.086(2)$\%&$0.285(1)$\%&$0.52(1)$\%\\
         $\lambda_2^\text{slow}\,[\tau_K^{-1}]$ & $2.903(3)\times10^{-4}$ & $5.468(5)\times10^{-4}$&$9.498(9)\times10^{-4}$\\
       $\lambda_2^\text{fast}\,[\tau_K^{-1}]$&$2.2(2)\times10^{-1}$&$4.15(8)\times10^{-1}$&$8.2(2)\times10^{-1}$\\\hline
         $A_3^\text{fast}$&$0.031(9)$\%&$0.000(7)$\%&$0.569(2)$\%\\
         $\lambda_3^\text{slow}\,[\tau_K^{-1}]$ &$ 5.47(2)\times10^{-4}$&$  1.09(3)\times10^{-3}$&$1.61(5)\times10^{-3}$\\
           $\lambda_3^\text{fast}\,[\tau_K^{-1}]$&$4(9)$&-&$8.0(4)\times10^{-1}$
   
    \end{tabular}
    \caption{ For $\epsilon_r=\langle\epsilon\rangle$, $\epsilon_r=4\langle\epsilon\rangle$ and $\epsilon_r=9\langle\epsilon\rangle$ and a number density of $400\text{cm}^{-3}$, the table displays fit parameters for the survival functions in Fig.~\ref{fig:rate_400}. $\lambda_1^\text{slow}$ corresponds to the exponential decay rate of the first size. Here we only have one exponential as the first collision cannot correlate with a previous one. Otherwise, we fitted a superposition of a short and long-time process. $\lambda_i^\text{slow}$ corresponds to the long-time collision rate while $\lambda_i^\text{fast}$ corresponds to the rate of the additional short-time process. Here, $A_i^\text{fast}=1 -A_i^\text{slow}$ denotes the weight of the short-time process and quantifies the fraction of droplets colliding due to correlations.}
    \label{tab:fit_lucky_400}
\end{table}
 As one might expect, the collision rates $\lambda_n^\text{slow}$ decrease in good approximation linearly with the density. As the ratios between the values of Fig.~\ref{tab:fit_lucky_400} and the corresponding values resemble the density ratio of 0.4, see Tab.~\ref{tab:fit_ratio}. We use this proportionality in the construction of the toy model to consider also number densities of $200\text{cm}^{-3}$. 

\begin{table}[h]
    \centering
    \begin{tabular}{c||c|c|c}
        & $\epsilon_r=\langle\epsilon\rangle$ &$\epsilon_r= 4\langle\epsilon\rangle$ & $\epsilon_r=9\langle\epsilon\rangle$ \\
        \hline
        $\lambda_1^\text{slow,A}/\lambda_1^\text{slow,B}$ & $0.40$ & $0.40$ & $0.39$ \\
        $\lambda_2^\text{slow,A}/\lambda_2^\text{slow,B}$ & $0.40$ & $0.40$ & $0.37$ \\
        $\lambda_3^\text{slow,A}/\lambda_3^\text{slow,B}$ & $0.38$ & $0.46$ & $0.39$ \\
    \end{tabular}
    \caption{Ratios of $\lambda_n^\text{slow}$ from Table~\ref{tab:fit_lucky_400} (A) to Table~\ref{tab:fit} (B) for $n=1,2,3$ and different $\epsilon_r$.}
    \label{tab:fit_ratio}
\end{table}

\newpage
\section{Sensitivity analysis of the toy model}
\begin{figure}[h]
    \centering
  \includegraphics{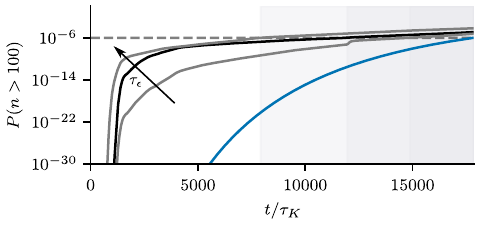}
    \caption{Probability of a droplet with more than a hundred collisions as a function of time. 
    The blue line corresponds to mean conditions, and the black line corresponds to the ensemble average over $10^5$ realizations of the toy model as presented in Fig.~5 of the main text. 
    The two dark gray lines correspond to the ensemble average over $10^5$ realizations of the toy model with approximately twice and half the correlation time for the local dissipation fluctuations than in the main text.
    The gray shaded areas mark the crossings of the $10^{-6}$ threshold.}
    \label{fig:corr_time}
\end{figure}

Here we change the correlation time $\tau_\epsilon$ of the random process within the toy model, which models the fluctuations of the local dissipation in time. By taking half and double the correlation time, we test how sensitive our results are in that regard.
In the main text, we reported a speed-up to bridge the size gap, i.e~of the one in a million fastest growing droplets to collide at least 100 times, of 33\% relative to mean conditions with a correlation time $\tau_\epsilon$ corresponding to about 19s with the assumed physical parameters, see Sec.~\ref{sec:para}. With almost half the correlation time, the speedup reduces to 17\%, and with roughly double the correlation time, it increases to 56\%, see Fig.~\ref{fig:corr_time}. This shows that the precise value of the speed-up is rather sensitive to the correlation time of the local dissipation fluctuations.
However, the qualitative result of a speed-up is also present for other plausible correlation times.

\end{document}